\input{aipcheck}

\documentclass[
    ,final            % use final for the camera ready runs
  ]
  {aipproc}

\layoutstyle{6x9}

\begin{document}

\title{Relativistic Shocks: Particle Acceleration, Magnetic Field Generation,
and Emission}

\author{K.-I. Nishikawa}{
  address={National Space Science and Technology Center,
  Huntsville, AL 35805 USA}
}

\author{P. Hardee}{
  address={Department of Physics and Astronomy,
  The University of Alabama,
  Tuscaloosa, AL 35487 USA}
}

\author{C. B. Hededal}{
  address={Niels Bohr Institute, Department of Astrophysics,
Juliane Maries Vej30, 2100 K\o benhavn \O, Denmark}
}

\author{G. Richardson}{
  address={Department of Mechanical and Aerospace Engineering
  University of Alabama in Huntsville Huntsville, AL 35899 USA}
}

\author{R. Preece}{
  address={Department of Physics,
  University of Alabama in Huntsville,
  Huntsville, AL 35899 and National Space Science and Technology Center,
  Huntsville, AL 35805 USA}
}

\author{H. Sol}{
  address={LUTH, Observatore de Paris-Meudon, 5 place Jules Jansen
  92195 Meudon Cedex, France}
}

\author{G. J. Fishman}{
  address={NASA-Marshall Space Flight Center, \\
National Space Science and Technology Center,
  Huntsville, AL 35805 USA}
}

\begin{abstract}
Shock acceleration is an ubiquitous phenomenon in astrophysical
plasmas.  Plasma waves and their associated instabilities (e.g.,
Buneman, Weibel and other two-stream instabilities) created in
collisionless shocks are responsible for particle (electron,
positron, and ion) acceleration. Using a 3-D relativistic
electromagnetic particle (REMP) code, we have investigated particle
acceleration associated with a relativistic  jet front propagating
into an ambient plasma with and without initial magnetic fields. We
find small differences in the results for no ambient and modest
ambient magnetic fields. Simulations show that the Weibel
instability created in the collisionless shock front accelerates jet
and ambient particles both perpendicular and parallel to the jet
propagation direction.  The non-linear fluctuation amplitudes of
densities, currents, electric, and magnetic fields in the
electron-positron shock are larger than those found in the
electron-ion shock at the same simulation time. This comes
from the fact that both electrons and positrons contribute to
generation of the Weibel instability.  While some Fermi acceleration
may occur at the jet front, the majority of electron and positron
acceleration takes place behind the jet front and cannot be
characterized as Fermi acceleration. The simulation results show
that the Weibel instability is responsible for generating and
amplifying nonuniform, small-scale (mainly transverse) magnetic 
fields which contribute
to the electron's (positron's) transverse deflection behind the jet
head.  This small scale magnetic field structure is appropriate to
the generation of ``jitter'' radiation from deflected electrons
(positrons) as opposed to synchrotron radiation.  The jitter
radiation has different properties than synchrotron radiation
calculated assuming a uniform magnetic field. The jitter radiation
resulting from small scale magnetic field structures may be
important for understanding the complex time structure and spectral
evolution observed in gamma-ray bursts and other astrophysical
sources containing relativistic jets and relativistic collisionless
shocks.

\end{abstract}

\maketitle

%%%%%%%%%%%%%%%%%%%%%%%%%%%%%%%%%%%%%%%%%%%%
%% MAINMATTER
%%%%%%%%%%%%%%%%%%%%%%%%%%%%%%%%%%%%%%%%%%%%

\section{Introduction}

Nonthermal radiation observed from astrophysical
systems containing relativistic jets and shocks, e.g., active
galactic nuclei (AGNs), gamma-ray bursts (GRBs), and Galactic
microquasar systems usually has power-law emission spectra. In most
of these systems, the emission is thought to be generated by
accelerated electrons through the synchrotron and/or inverse Compton
mechanisms. Radiation from these systems is observed from the radio
through the gamma-ray region. Radiation in optical and higher
frequencies typically requires particle acceleration in order to
counter radiative losses.  It has been proposed that the needed
particle acceleration occurs in shocks produced by differences in
flow speed.

Particle-in-cell (PIC) simulations can shed light on the physical
mechanism of particle acceleration that occurs in the complicated
dynamics within relativistic shocks.  Recent PIC simulations using
injected relativistic electron-ion jets show that acceleration
occurs within the downstream jet, rather than by the scattering of
particles back and forth across the shock as in Fermi acceleration
\cite{fred02,fred04,nis03,nishi04}, and \citet{silva03}
have presented
simulations of the collision of two inter-penetrating
electron-positron plasma shells as a model of an astrophysical
collisionless shock. In the electron-positron simulations performed
with counter-streaming jets \cite{silva03}, shock dynamics
involving the propagating jet head (where Fermi acceleration may
take place) was not investigated. In general, these independent
simulations have confirmed that relativistic jets excite the Weibel
instability \cite{weib59}.  The Weibel instability generates current
filaments and associated magnetic fields \cite{medv99},
and accelerates
electrons \cite{silva03,fred02,fred04,nis03,hede04}.

In this paper we present new simulation results of particle
acceleration and magnetic field generation for relativistic
electron-positron and electron-ion shocks using 3-D relativistic
electromagnetic particle-in-cell (REMP) simulations. In our new
simulations, electron-positron and electron-ion relativistic jets
with Lorentz factor, $\gamma = 5$ (corresponds to 2.5 MeV) is
injected into electron-positron and electron-ion plasmas in order to
study the dynamics of a relativistic collisionless shock both with
and without an initial ambient magnetic field.

\section{Simulation Setup and results}

Four simulations were performed using an $85 \times
85 \times 320$ grid with a total of 180 million particles (27
particles$/$cell$/$species for the ambient plasma) and an electron
skin depth, $\lambda_{\rm ce} = c/\omega_{\rm pe} = 9.6\Delta$,
where $\omega_{\rm pe} = (4\pi e^{2}n_{\rm e}/m_{\rm e})^{1/2}$ is
the electron plasma frequency and $\Delta$ is the grid size.  In all
simulations jets are injected at $z = 25\Delta$ in the positive $z$
direction. In all simulations radiating boundary conditions were
used on the planes at $z =0,z_{\rm max}$. Periodic boundary
conditions were used on all other boundaries \cite{bun93}. 

In two simulations an electron-positron jet is injected into a
magnetized and unmagnetized electron-positron ambient plasma and in two
simulations an electron-ion jet is injected into a magnetized and
unmagnetized electron-ion ambient plasma.  The choice of parameters and
simulations allows comparison with previous simulations
\cite{silva03,fred02,fred04,hede04,nis03,nishi04}.

The electron number density of the jet is
$0.741n_{\rm b}$, where $n_{\rm b}$ is the density of ambient
(background) electrons. The average jet velocity is $v_{\rm j} =
0.9798c$, and the corresponding Lorentz factor is 5. The jet is cold
($v^{\rm e}_{\rm j, th} = v^{\rm p}_{\rm j, th} = 0.01c$
and $v^{\rm i}_{\rm j, th} = 0.0022c$ in the laboratory frame).  The
ambient and jet electron-positron plasma has mass ratio $m_{\rm
p}/m_{\rm e} \equiv m_{\rm e^+}/m_{\rm e^-} = 1$
($m_{\rm i}/m_{\rm e} = 20$).  The electron and ion
thermal velocities in the ambient plasma are $v^{\rm e}_{\rm th} = 0.1c$
and $v^{\rm i}_{\rm th} = 0.022$, respectively, where
$c$ is the speed of light.
The time step $t = 0.013/\omega_{\rm pe}$,
the ratio $\omega_{\rm pe}/\Omega_{\rm e} = 11.5$, and the Alfv\'en
speed (for electrons) $v_{\rm Ae} \equiv (\Omega_{\rm e}/\omega_{\rm
pe})c = 8.66\times 10^{-2}c$. With the speed of an Alfv\'en wave given by
$v_{\rm A} = [V_{\rm A}^{2}/(1 + V_{\rm A}^{2}/c^{2})]^{1/2}
= 6.10 \times 10^{-2}c$
where $V_{\rm A} \equiv [B^{2}/4\pi (n_{\rm e} m_{\rm e}
+ n_{\rm p} m_{\rm p})]^{1/2} = 6.12 \times 10^{-2}c$, the Alfv\'en Mach number
$M_{\rm A} \equiv v_{\rm j}/v_{\rm A} = 16.0$.
With a magnetosonic speed $v_{\rm ms} \equiv (v_{\rm
th}^{2} + v_{\rm A}^{2})^{1/2} = 0.132c$ the magnetosonic Mach number $M_{\rm
ms} \equiv v_{\rm j}/v_{\rm ms} = 7.406$. At least approximately the
appropriate relativistic Mach numbers multiply these values by the
Lorentz factor.  Thus, in an MHD approximation we are dealing with a
high Mach number shock with $\gamma M >> 1$. The gyroradius of
ambient electrons and positrons with $v_{\perp} = v_{\rm th} = 0.1c$
is $11.1\Delta = 1.154\lambda_{\rm ce}$ (for ambient ions: $49.6\Delta =
5.16\lambda_{\rm ce}$).  All the Mach numbers with electron-ion jets
are approximately increased by $\sqrt{m_{\rm i}/m_{\rm e}} = \sqrt{20}
= 4.47$.

%Fig. 1
\begin{figure}[ht]
\includegraphics[height=.62\textheight]{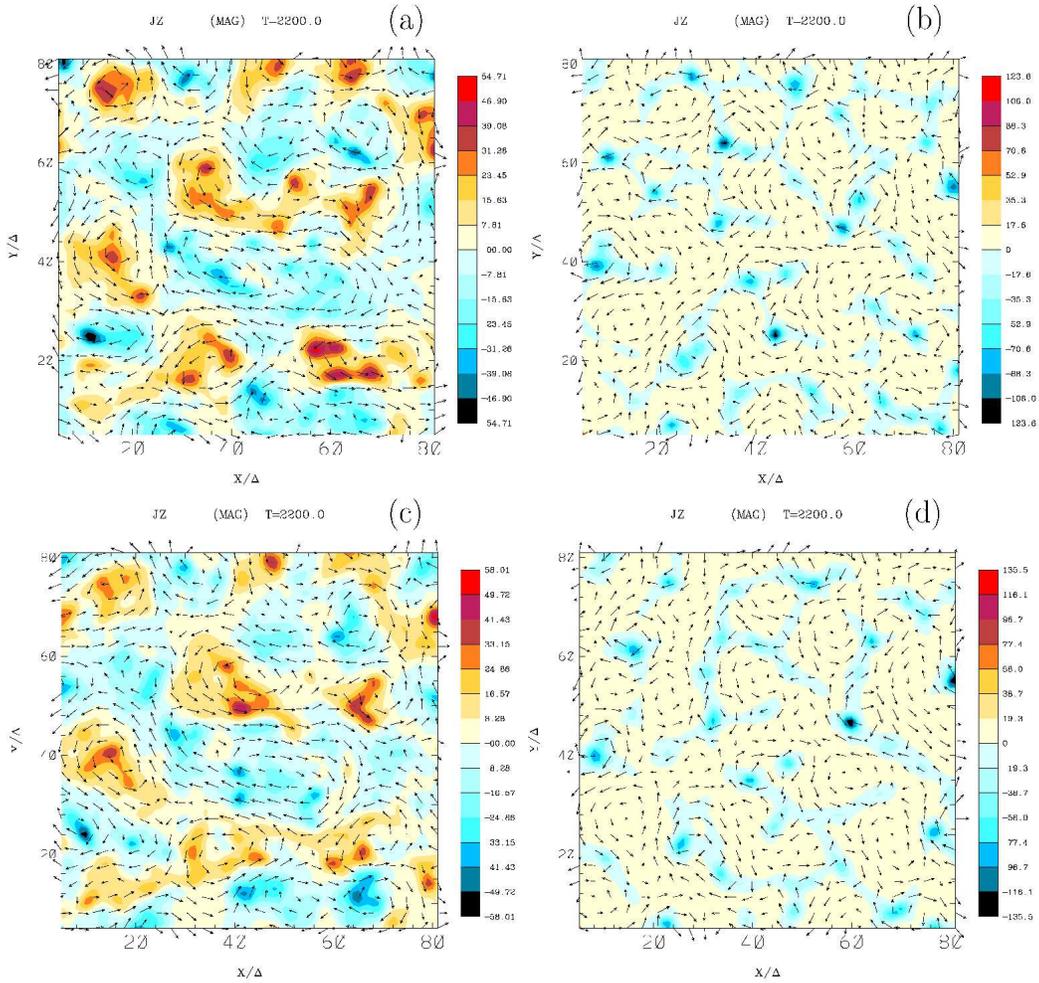}
\caption{2D images in the $x - y$ plane at $z =
230\Delta$ for a flat jet injected into a unmagnetized ((a) and (b))
and magnetized ((c) and (d)) ambient medium shown at $t =
28.8/\omega_{\rm pe}$. Colors indicate the $z$-component of the
current density ($J_{\rm z}$) ((a) and (c): electron-positron, (b)
and (d): electron-ion) (peaks: (a) $\pm 54.7$, (b) $-123.6$, (c) $\pm 58.0$, 
(d) $-135.5$ with $B_{\rm x}, B_{\rm y}$ indicated by the arrows.}
\end{figure}

\begin{figure}[ht]
\includegraphics[height=.62\textheight]{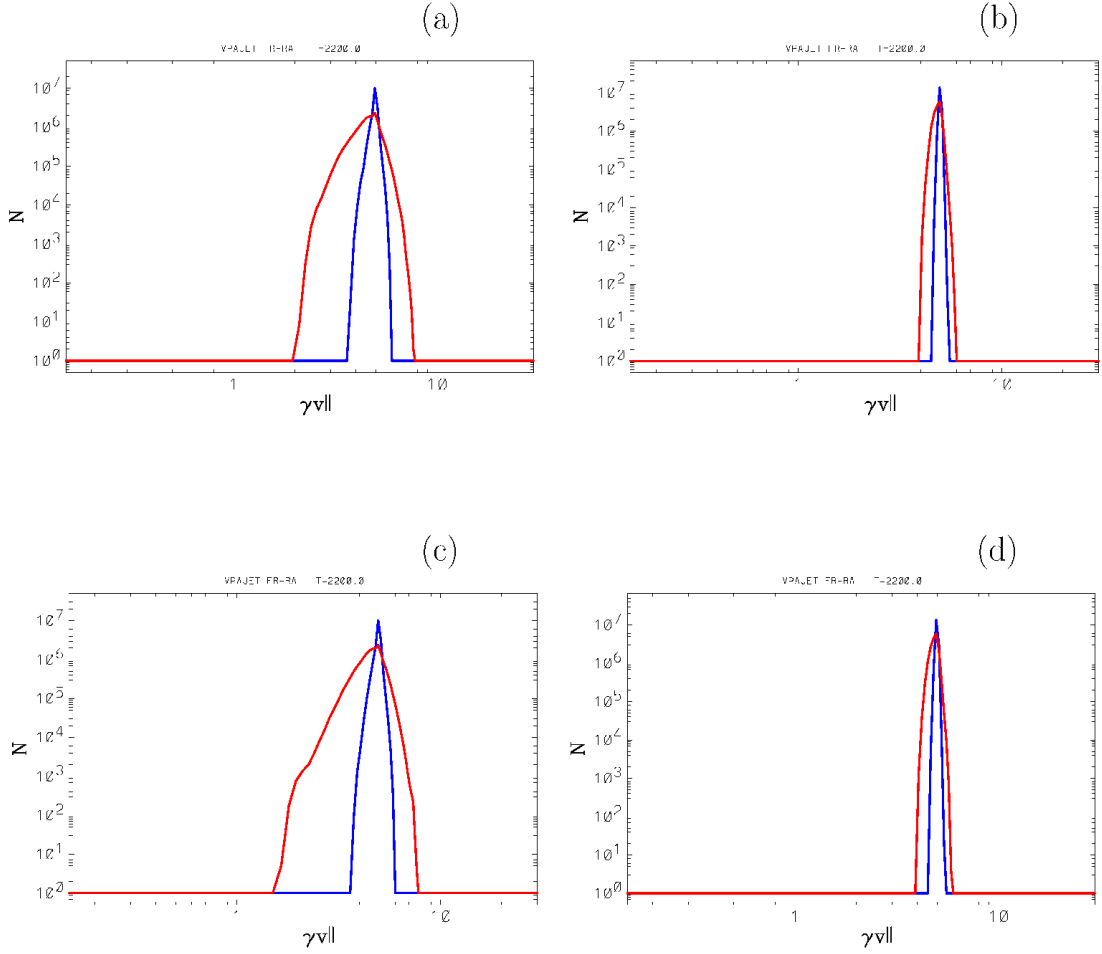}
\caption{Velocity distributions of jet electrons four cases as
Fig. 1 at $t = 28.8/\omega_{\rm pe}$ ((a): electro-positron, unmagnetized,
(b) electron-ion, unmagnetized, (c)electro-positron, magnetized, and
(d) electron-ion, magnetized). The blue and red curves show the distributions of
injected and shocked jet electrons. Jet electrons are binned as a function 
of $\gamma v_{\parallel}$, where $\gamma = (1 -(v^{2}_{\parallel} 
+v^{2}_{\perp})/c^{2})^{-1/2}$.}
\end{figure}

Current filaments resulting from development of the Weibel
instability behind the jet front (at $z = 230\Delta$) are shown 
in Figure 1 at time $t =
28.8/\omega_{\rm pe}$ for four different cases. The upper two panels (a and b)
show for unmagnetized ambient plasmas and the lower two panels (c and d) are
for magnetized ambient plasmas. In cases (a and c) an electron-positron jet
is injected in an electron-positron ambient plasma. An electron-ion jet
is injected in an electron-ion ambient plasma (b an d). The maximum
values of $J_{\rm z}$ are (a) $\pm 54.7$, (b)$ -123.6$, (c) $\pm 58.0$, 
and (d) $-135.5$. The electro-positron jets show electron (negative) and 
positron (positive) current
filamentations, since both species contribute to the Weibel instability.
On the other hind, at this simulation time mainly electron jets generate 
(negative) current filamentations as shown in Figs. 1 b and 1d.
The effect of weak ambient magnetic fields increase the maximum values of
current filamentations slightly.

The electrons are deflected by the transverse magnetic fields
($B_{\rm x}, B_{\rm y}$) via the Lorentz force: $-e({\bf v} \times
{\bf B})$, generated by current filaments ($J_{\rm z}$), which in
turn enhance the transverse magnetic fields \cite{weib59,medv99}. 
The complicated filamented structures resulting from
the Weibel instability have diameters on the order of the electron
skin depth ($\lambda_{\rm ce} = 9.6\Delta$). This is in good
agreement with the prediction of $\lambda \approx
2^{1/4}c\gamma_{\rm th}^{1/2}/\omega_{\rm pe} \approx
1.188\lambda_{\rm ce} = 11.4\Delta$ \cite{medv99}. Here,
$\gamma_{\rm th} \sim 1$ is a thermal Lorentz factor. 
However, in the electron-positron jets
the current filaments are coalesced in the transverse direction,
and this shows the nonlinear evolution. The longitudinal current
($J_{\rm z}$) in the electron-positron jets (a and c)
shows significantly more transverse variation than in the
electron-ion jets (b and d).

The acceleration of electrons has been reported in previous work
\cite{silva03,fred02,fred04,nis03,nishi04,hede04}.
Figure 2 shows that the cold jet electrons are accelerated and
decelerated. As expected, at this time jet electrons in the electron-positron
jets are thermalized more strongly than in the electron-ion jets. 
The blue curves in Figs. 2b and 2d is close to the initial distribution 
of injected jet electrons.   
We also see that the kinetic energy (parallel velocity $v_{\parallel}
\approx v_{\rm j}$) of the jet electrons is transferred to the
perpendicular velocity via the electric and magnetic fields
generated by the Weibel instability \cite{nishi04}. The strongest transverse
acceleration of jet electrons accompanies the strongest
deceleration of electron flow and occurs between
$z/\Delta = 210 - 240$.  The transverse acceleration in the electron-positron
jets is over four times that in the electron-ion simulations.
The strongest acceleration takes place around the maximum amplitude of
perturbations due to the Weibel instability at $z/\Delta \sim 220$
as seen in Figs 1a and 1c.

\section{Summary and Discussion}

We have performed self-consistent,
three-dimensional relativistic particle simulations of relativistic
electron-positron and electron-ion jets propagating into magnetized
and unmagnetized electron-positron and electron-ion ambient plasmas.
The main acceleration of electrons takes place in the downstream
region. Processes in the relativistic collisionless shock are
dominated by structures produced by the Weibel instability.  This
instability is excited in the downstream region behind the jet head,
where electron density perturbations lead to the formation of
current filaments \cite{nishi04}. The nonuniform electric field and 
magnetic field structures associated with these current filaments 
thermalize the jet electrons and positrons, while accelerating the 
ambient electrons and positrons, and accelerating (heating) the jet 
and ambient electrons and positrons in the transverse direction.

Other simulations with different skin depths and plasma frequencies
show that the growth and structure of current filaments generated by
the Weibel instability scale with the plasma frequency and the skin
depth \cite{nishi04}.  An additional simulation in which an
electron-ion jet is injected into a ambient plasma with perpendicular
magnetic field shows magnetic reconnection due to the generation of an
antiparallel magnetic field generated by bending of jet electron
trajectories \cite{hede04a}.

This small scale magnetic field structure generated by the Weibel
instability is appropriate to the generation of ``jitter'' radiation 
from deflected electrons (positrons) \cite{medv00}. 
The jitter radiation
resulting from small scale magnetic field structures needs to be 
calculated from the trajectories of electrons and positrons (ions)
which requires extensive computational resources. 
This investigation will provide an idea for understanding the complex 
time structure and spectral evolution
observed in gamma-ray bursts and other astrophysical sources containing
relativistic jets and relativistic collisionless shocks.

The fundamental characteristics of relativistic shocks are essential
for a proper understanding of the prompt gamma-ray and afterglow
emission in gamma-ray bursts, and also to an understanding of the
particle reacceleration processes and emission from the shocked regions
in relativistic AGN jets.  Since the shock dynamics is complex and
subtle, more comprehensive studies using larger systems 
are required to better understand
the acceleration of electrons, the generation of magnetic fields and
the associated emission. This further study will provide the insight
into basic relativistic collisionless shock characteristics needed to
provide a firm physical basis for modeling the emission from shocks in
relativistic flows.

\begin{theacknowledgments}

K. Nishikawa is a NRC Senior Research Fellow
at NASA Marshall Space Flight Center. This research (K.N.) is
partially supported by the National Science Foundation awards ATM
9730230, ATM-9870072, ATM-0100997, and INT-9981508. P. Hardee
acknowledges partial support by a National Space Science and
Technology (NSSTC/NASA) award.  The simulations have been performed
on ORIGIN 2000 and IBM p690 (Copper) at the National Center for
Supercomputing Applications (NCSA) which is supported by the
National Science Foundation.

\end{theacknowledgments}

%%%%%%%%%%%%%%%%%%%%%%%%%%%%%%%%%%%%%%%%%%%%%%%%
%% You may have to change the BibTeX style below, depending on your
%% setup or preferences.
%%
%% If the bibliography is produced without BibTeX comment out the
%% following lines and see the aipguide.pdf for further information.
%%
%% For The AIP proceedings layouts use either
%%%%%%%%%%%%%%%%%%%%%%%%%%%%%%%%%%%%%%%%%%%%

\bibliographystyle{aipproc}   % if natbib is available
%\bibliographystyle{aipprocl} % if natbib is missing

%%%%%%%%%%%%%%%%%%%%%%%%%%%%%%%%%%%%%%%%%%%
%% You probably want to use your own bibtex database here
%%%%%%%%%%%%%%%%%%%%%%%%%%%%%%%%%%%%%%%%%%%
%\bibliography{hegra04}

%%%%%%%%%%%%%%%%%%%%%%%%%%%%%%%%%%%%%%%%%%%
%% Just a reminder that you may have to run bibtex
%% All of it up to \end{document} can be removed
%% if you don't like the warning.
%%%%%%%%%%%%%%%%%%%%%%%%%%%%%%%%%%%%%%%%%%%

\end{document}